\documentclass[final,3p,times,twocolumn]{elsarticle}
\usepackage{amssymb, ams math}
\journal{Physics Letters A}
\biboptions{sort&compress}

\usepackage{graphicx}
\usepackage{lineno,todonotes}

\newcommand{\KY}{\mathrm{KY}}
\newcommand{\Trans}{\mathcal{T}}
\newcommand{\argmax}[1]{\underset{#1}{\mathrm{arg\,max\ }}}

\begin{document}
\begin{frontmatter}

\title{Analysing spatially extended high-dimensional dynamics by recurrence plots}

\author[1]{Norbert Marwan}
\ead{marwan@pik-potsdam.de}
\author[1,2,3]{J\"urgen Kurths}
\author[4]{Saskia Foerster}
\address[1]{Potsdam Institute for Climate Impact Research, 14412 Potsdam, Germany}
\address[2]{Humboldt Universit\"at zu Berlin, Institut f\"ur Physik, Germany}
\address[3]{Nizhny Novgorod State University, Department of Control Theory, Nizhny Novgorod, Russia}
\address[4]{GFZ German Research Centre for Geosciences, Section 1.4 Remote Sensing, Telegrafenberg, 14473 Potsdam,
 Germany}

\date{\today}

\begin{abstract}
Recurrence plot based measures of complexity are capable tools for characterizing complex dynamics.
In this
letter we show the potential of selected recurrence plot measures for the investigation
of even high-dimensional dynamics. We apply this method on spatially extended chaos, such 
as derived from the Lorenz96 model and show that the recurrence plot based measures
can qualitatively characterize typical dynamical properties such as chaotic or periodic dynamics.
Moreover, we demonstrate its power by analysing satellite image time series of vegetation 
cover with contrasting dynamics as a spatially extended and potentially high-dimensional example from the real world.
\end{abstract}
\end{frontmatter}

\section{Introduction}
The recurrence plot (RP) is a modern and versatile tool for the study of the complex behavior of dynamical systems \cite{eckmann87,marwan2008epjst}.
It represents time points of recurring states even of high-dimensional phase space trajectories.
Quantitative extensions, such as recurrence quantification analysis and recurrence networks, 
enable the investigation of dynamical transitions and regime changes, the quantitative characterization of 
the dynamics, or the detection of phase synchronization \cite{marwan2007,donner2011epjb,webber2015}. 
As proven by several examples, the RP based quantities work quite well even with short time series 
(e.g.,\cite{marwan2009b,litak2010,kopacek2010,guimaraes2010}).
The practical and powerful use of RP based methods has been demonstrated by their
growing and interdisciplinary application, such as for cardiovascular health diagnosis, behavioral, cognitive and
neurological studies, studying fluid dynamics and plasma, analysing optical effects, 
material health monitoring, palaeoclimate regime change detection, 
etc.~\cite{ramirez2013b,richardson2005,eneyew2013,guimaraes2010,tahmasebpour2013b,krese2011b,konvalinka2011,donges2011d}. 
In general, such studies have so far been restricted to rather low-dimensional systems.
However, when studying the 
complex behavior of real world systems, we often end up with extended complex systems, and the
question arises whether the RP based tools can be applied on high-dimensional 
systems, such as exhibiting high-dimensional chaos. So far, the ability of RP based methods
for studying high-dimensional dynamics has not yet been demonstrated, although it was already
used to investigate spatial recurrences \cite{vasconcelos2006,marwan2007pla,prado2014} and 
spatio-temporal chaos in turbulence and a reaction-diffusion system \cite{guimaraes2008,mocenni2010}.
Moreover, the classic characterization of complex dynamics by using, e.g., entropy \cite{eckmann85}, 
correlation dimension \cite{grassberger83}, and Lyapunov exponents
requires very long time-series \cite{eckmann1992} or the knowledge of the differential equations of the system which are in 
practical examples not known. The study of extended spatio-temporal dynamics
is even more challenging because of the large degrees of freedom.

\begin{figure*}[htbp]
\centering \includegraphics[width=.7\textwidth]{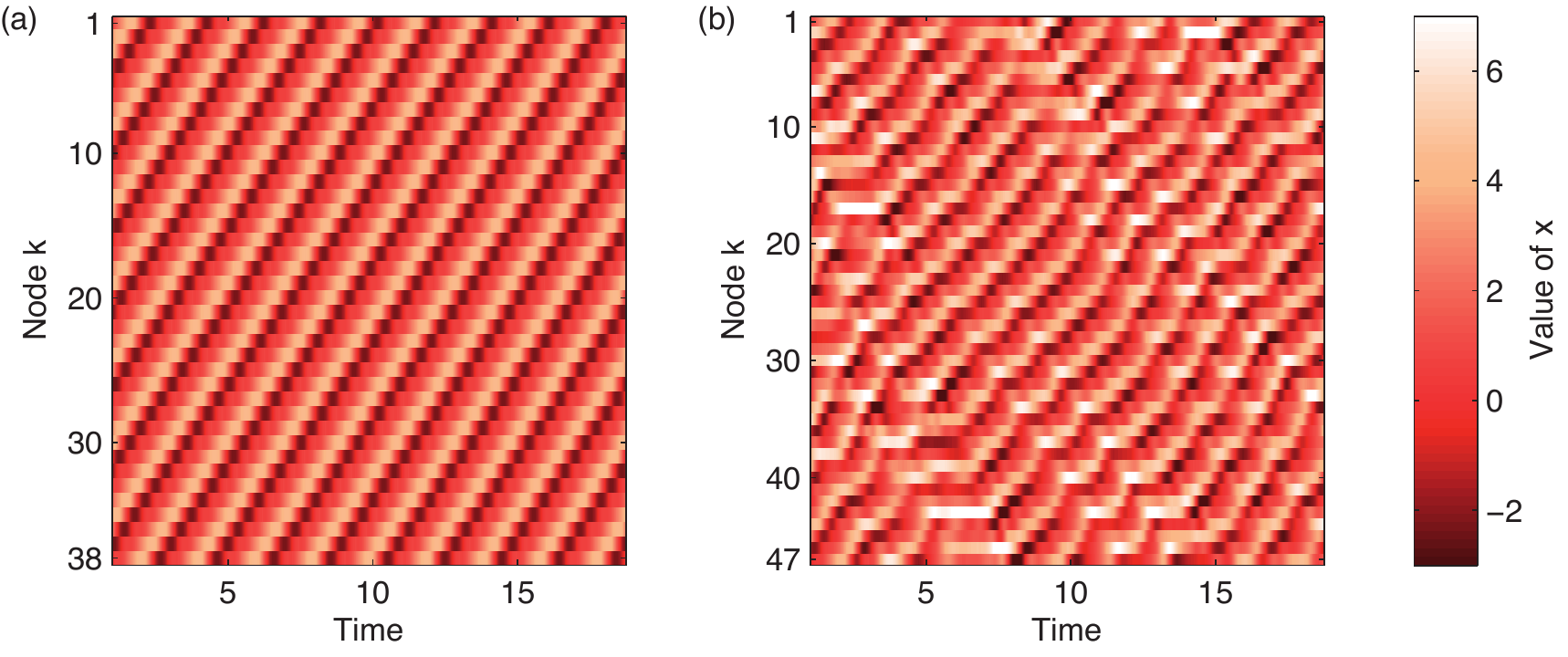} 
\caption{Space-time representation of $x_k(t)$ of the Lorenz96 system, Eq.~(\ref{eq_lorenz96}), for $f=5$ and system size
of (a) $N=38$ and (b) $N=47$, showing periodic and chaotic dynamics.}\label{fig_data}
\end{figure*}

In this letter we demonstrate the potential of RP based measures of complexity for
identifying hardly accessible extended spatio-temporal dynamics and for
characterizing high-dimensional chaos. We will use the Lorenz96 model \cite{lorenz96,lorenz1998,pazo2008}
which is a paradigmatic system for extended complex spatio-temporal chaotic dynamics and 
was systematically studied by
Karimi et al.~\cite{karimi2010} and apply the method on an example of a satellite time series imagery.

\section{The Lorenz96 model}
The Lorenz96 model is a conceptual time-continuous linear lattice model that was
developed to demonstrate fundamental aspects of weather predictability \cite{lorenz96}:
\begin{equation}\label{eq_lorenz96}
\frac{dx_k}{dt} = (x_{k+1} - x_{k-2})x_{k-1} - x_{k-1} + f
\end{equation}
for $k = 1, \ldots, N$, with a constant external forcing $f$, and
with periodic boundary conditions $x_{N+1} = x_1$.
Depending on the system size $N$ and the forcing $f$, the 
dynamics on the lattice can be periodic or chaotic and
can exhibit a high dimensionality \cite{karimi2010}. Therefore, this model is 
very appropriate for our study.

For integrating Eq.~(\ref{eq_lorenz96}) we use a
Runge-Kutta integration of 4th order with time step 
$\delta t = 1/64$. In order to remove transients, we
neglect the first 10,000 values from each $x_k(t)$.
In the numerical experiments discussed below, we
will use 20 slightly varying initial conditions for each selected 
setting of $N$ and $f$.

In our study we consider $f=5$ (as used by Karimi et al.~\cite{karimi2010}). 
Then, for example, for $N=38$, we find periodic dynamics, but for $N=47$,
the dynamics is chaotic (Fig.~\ref{fig_data}).

\begin{figure}[htbp]
\centering \includegraphics[width=.9\columnwidth]{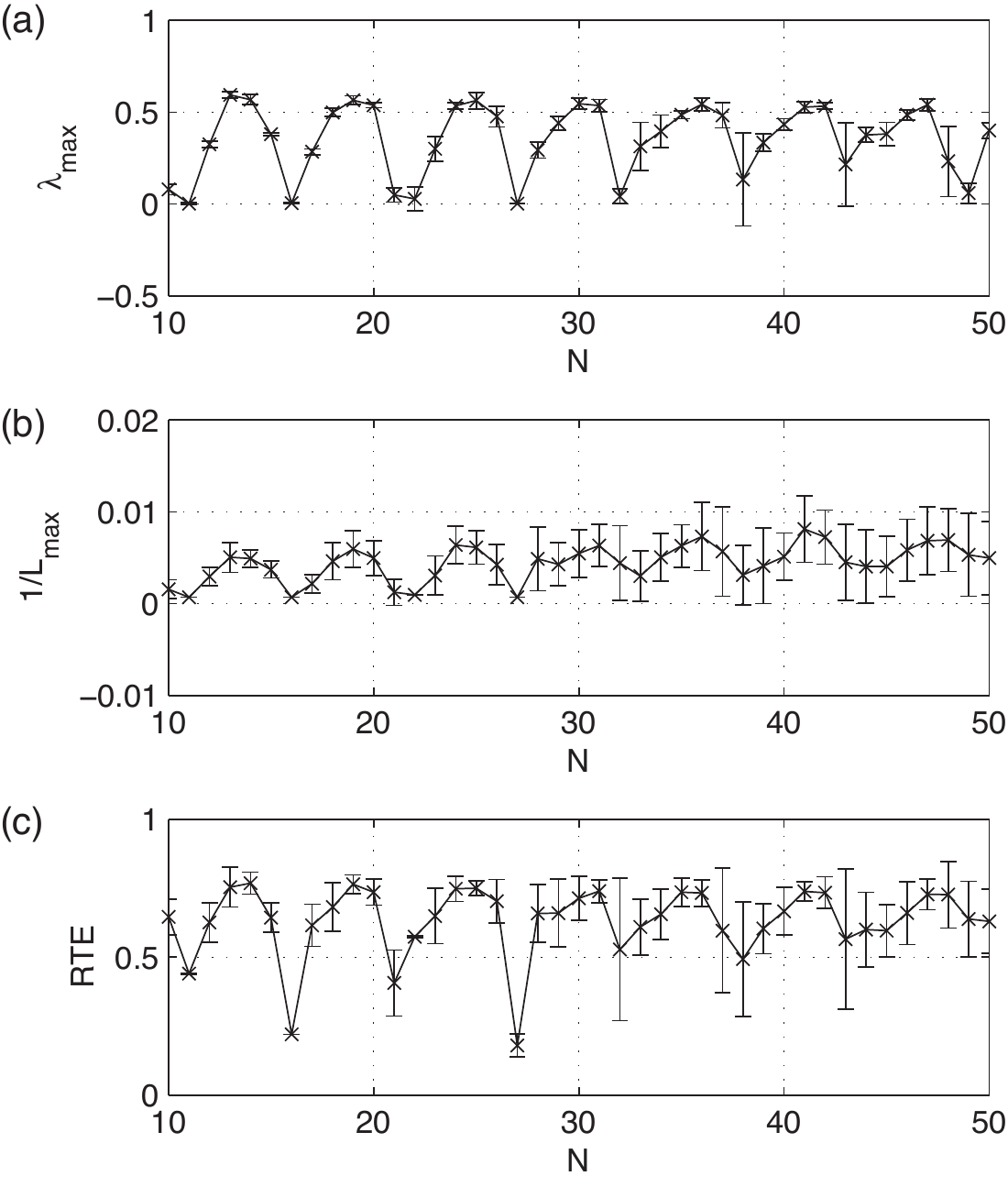} 
\caption{(a) Maximal Lyapunov exponent $\lambda_{\max}$ for the Lorenz96 system, Eq.~(\ref{eq_lorenz96}), with different system size $N$.
The RP based measures (b) $1/L_{\max}$ and (c) RTE reveal a similar variation
with the $N$ as $\lambda_{\max}$. Averaged values for 20 different initial
states are presented. The standard deviations of the measures for the different
initial conditions are presented by the error bars.}\label{fig_lyap}
\end{figure}

\begin{figure}[htbp]
\centering \includegraphics[width=.9\columnwidth]{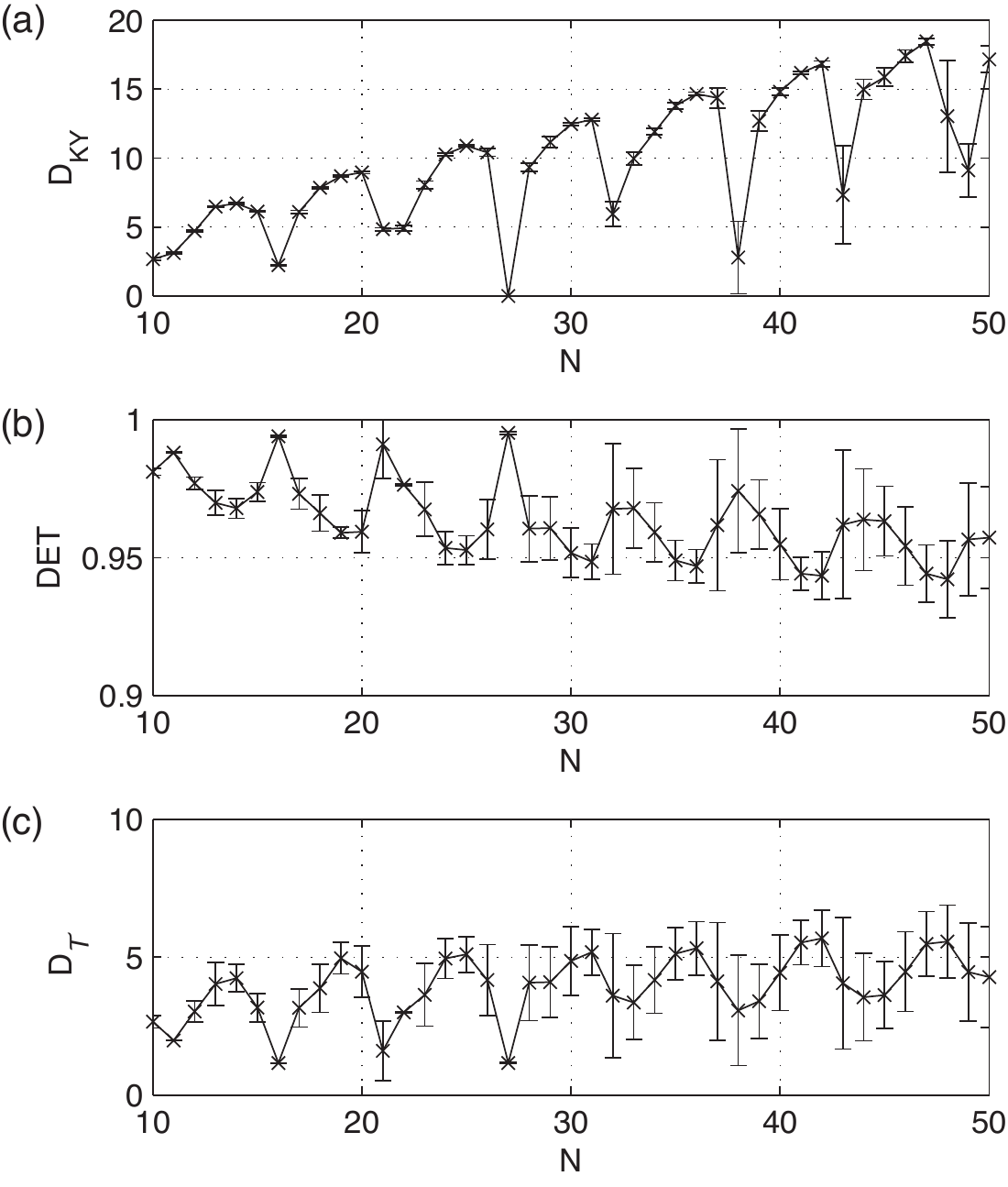} 
\caption{(a) Kaplan-Yorke dimension $D_{\KY}$ for the Lorenz96 system, Eq.~(\ref{eq_lorenz96}), with different system size $N$.
The RP based measures (b) DET and (c) $D_{\Trans}$ reveal a similar variation
with the $N$ as $D_{\KY}$. Averaged values for 20 different initial
states are presented. The standard deviations of the measures for the different
initial conditions are presented by the error bars.}\label{fig_dim}
\end{figure}

The change of the dynamical regimes with system size $N$ can be measured by the maximal
Lyapunov exponent $\lambda_{\max}$ and the Kaplan-Yorke dimension $D_{\KY}$.
Here we compute the Lyapunov spectrum from the set of $N$ differential equations
by linearizing the corresponding evolution and using a Gram-Schmidt Orthonormalization 
scheme \cite{christiansen1997,ramasubramanian2000}. For stable results, we 
integrate 200,000 iterations. The Kaplan-Yorke dimension $D_{\KY}$ can then be derived from
the $N$ (ordered) Lyapunov exponents by the Kaplan-Yorke algorithm
\begin{equation}\label{eq_D_KY}
D_{\KY} = K + \sum_{i=1}^K \frac{\lambda_i}{|\lambda_{K+1}|},
\end{equation}
where $K$ is the largest number of the first largest Lyapunov exponents with 
$\sum_{i=1}^K \lambda_i \ge 0$ \cite{eckmann85}. 
Increasing the system size from $N=10$ to $N=50$
reveals a periodic alternation between periodic and chaotic dynamics by periodic
variations of $\lambda_{\max}$ (Fig.~\ref{fig_lyap}a). The dimension of the system dynamics
as measured by $D_{\KY}$ is increasing by trend (Fig.~\ref{fig_dim}a). The calculation
of $\lambda$ and $D_{\KY}$ is expensive for such systems with large degrees of freedom.
Moreover, for accurate values we need very long time series (here, even for $N=200,000$ we find
some spread in the results of $\lambda_{\max}$ and $D_{\KY}$).

\section{Recurrence plot analysis}

\begin{figure}[htbp]
\centering \includegraphics[width=\columnwidth]{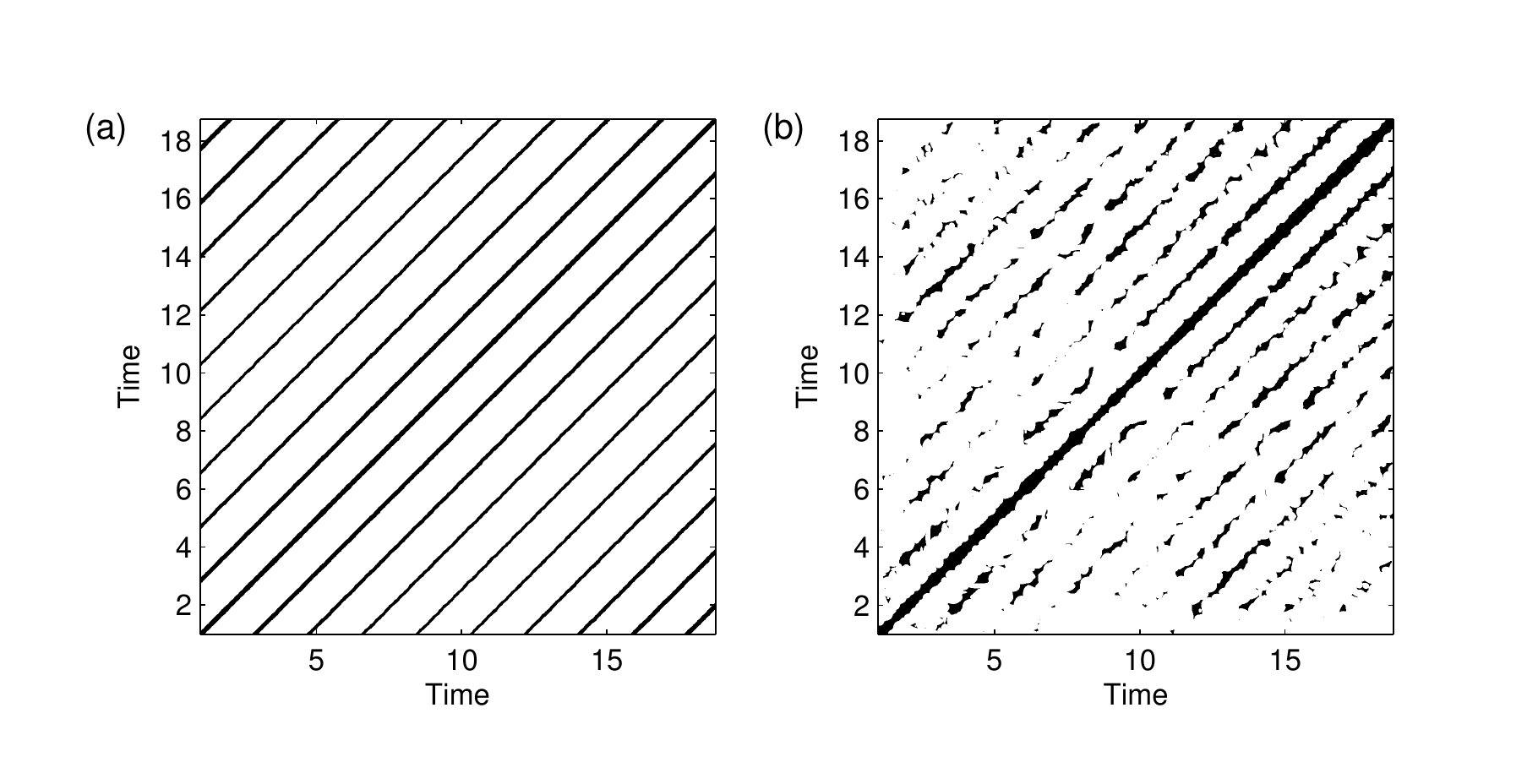} 
\caption{Recurrence plots of the Lorenz96 system $x_k(t)$ 
for $f=5$ and system size
of (a) $N=38$ and (b) $N=47$, showing periodic and chaotic 
dynamics.}\label{fig_rp_lor96}
\end{figure}

RP quantification may be suitable for a simpler estimation of the dynamical properties. An RP 
$R_{i,j} = \Theta(\varepsilon - \| \vec{x}_i - \vec{x}_j\|)$ is a binary matrix $\mathbf{R}$ representing
the time points $j$ when a state $\vec{x}_i$ at time $i$ recurs \cite{marwan2007}
(Fig.~\ref{fig_rp_lor96}). 
The recurrence
criterion is usually defined as a spatial distance between two states $\vec{x}_i$ and
$\vec{x}_j$ is falling below a threshold $\varepsilon$. Besides the ability to discuss
the visual aspect of an RP, several quantification approaches are based on this matrix. 
The diagonal line structures in an RP correspond to periods of parallel evolution of
two segments of the phase space trajectory. The scaling of the length distribution of such
lines is related to the $K_2$ entropy. A good proxy for this is measuring the inverse of
the length of the longest diagonal line $1/L_{\max}$, with 
\begin{equation}\label{eq_det}
L_{\max} = \argmax{l} H_D(l),
\end{equation}
and $l$ the length of the diagonal lines, and
$H_{\text{D}}(l)$ the length distribution of diagonal lines in $\mathbf{R}$ \cite{marwan2007}.

Based on a heuristic approach,
the fraction of recurrence points that form such diagonal lines is a qualitative measure
of predictability, called {\it determinism (DET)} \cite{marwan2007},
\begin{equation}\label{eq_det}
DET=\frac{\sum_{l=2}^N l\, H_{\text{D}}(l)}{\sum_{i,j=1}^N R_{i,j}}.
\end{equation}
Systems possessing deterministic dynamics are characterized by diagonal lines 
indicating repeating recurrences within a state (and, hence, higher $DET$ values).

\begin{figure*}[hbtp]
\centering \includegraphics[width=\textwidth]{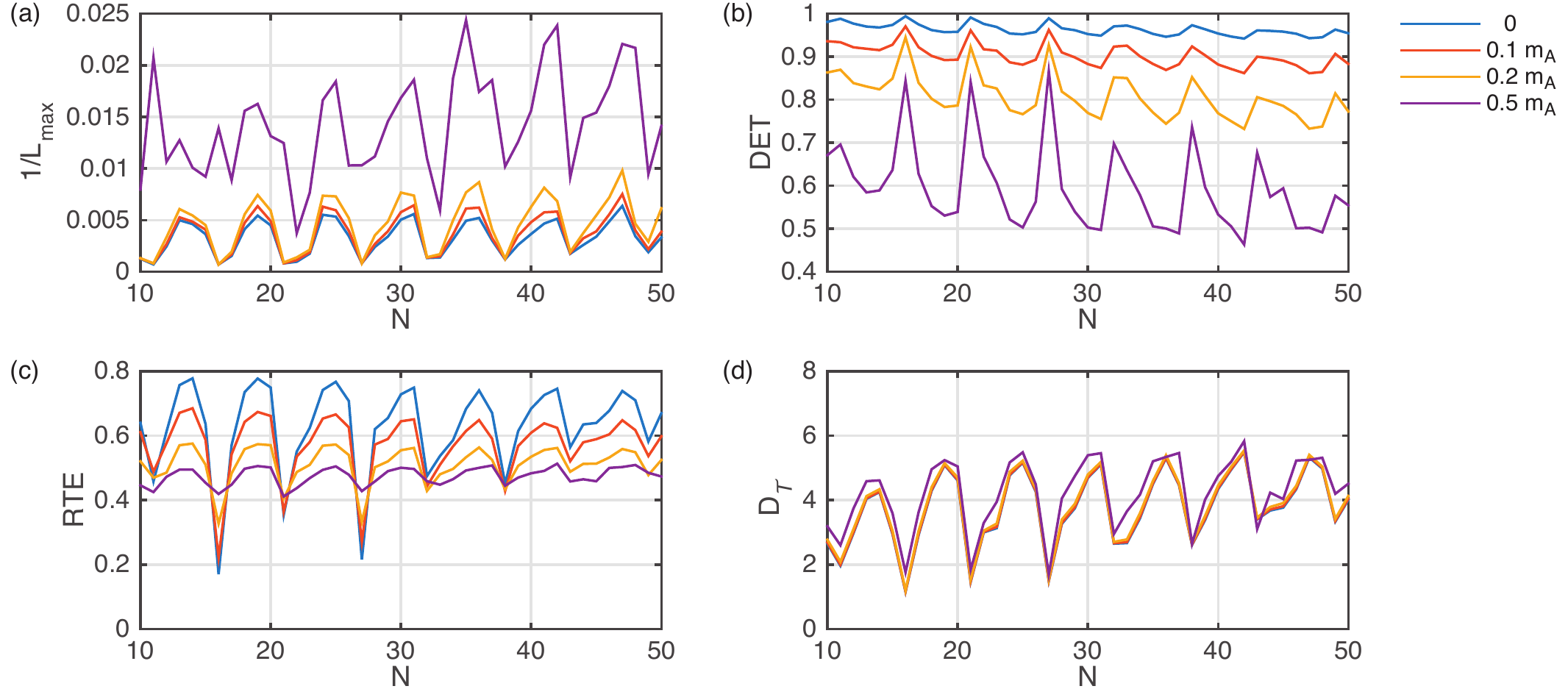} 
\caption{Influence of noise on the RQA results. Observational, normally distributed noise is added to the time series
of the Lorenz96 system. The standard deviation of the noise is relative to the averaged mean amplitude $m_A$ of the
time series.}\label{fig_noise}
\end{figure*}

The vertical empty space between two recurrence points in the RP
correspond to Poincar\'e recurrence times, i.e., the distance $v$ between recurrence points in a 
column of $\mathbf{R}$ \cite{ngamga2012}.
From the distribution $H_{\text{V}}(v)$
we can derive the {\it recurrence time entropy (RTE)}, also called {\it recurrence period density entropy} \cite{little2007}
\begin{equation}\label{eq_rpde}
RTE=-\frac{1}{\ln{V_{\max}}} \sum_{v=1}^{V_{\max}}
   H_{\text{V}}(v) \ln{H_{\text{V}}(v)}.
\end{equation}
This measure quantifies the extent of recurrences and is related to the Pesin
dimension \cite{anishchenko2013}.

In the last years, the similarity of the binary, squared matrix 
$\mathbf{R}$  with the adjacency matrix of an unweighted, undirected complex network was
used to apply complex network measures on recurrence plots 
in order to quantify the geometrical properties of the system's attractor 
encoded in the RP \cite{marwan2009b}. For example, the 
{\it transitivity coefficient} ($\mathcal{T}$)
\begin{equation}\label{eq_trans}
\mathcal{T} =  \frac{\sum_{i,j,k=1}^N R_{j,k}R_{i,j}R_{i,k}}{\sum_{i,j,k=1}^N R_{i,j}R_{i,k} (1-\delta_{j,k})}.
\end{equation}
allows the differentiation of periodic and chaotic dynamics \cite{zou2010}. 
Moreover, $\mathcal{T}$ can be used to define a novel dimensionality measure, the
{\it transitivity dimension} ($D_{\mathcal{T}} $)
 \begin{equation}\label{eq_transdim}
D_{\mathcal{T}} = \frac{\log(\mathcal{T})}{log(3/4)},
\end{equation}
allowing the calculation of the dimension without
explicit consideration of scaling behaviors.
Using the RP, the correlation dimension $D_2$ can also be derived \cite{grassberger83c}.
However, the advantage of $D_{\Trans}$ is that it results directly from the
RP without analysing any scaling behavior depending on the recurrence threshold $\varepsilon$.

Although still rather novel, such recurrence quantification is meanwhile widely
accepted and applied in different disciplines to study diverse problems. For more details
on this methodology we refer to \cite{marwan2007,marwan2011,marwan2014,webber2015}.

\section{Recurrence analysis of spatially extended chaos}
For the application of the RP approach to spatially extended high-dimensional data such
as from the Lorenz96 model, we consider each variable as one component
of the phase space representation: $\vec x(t) = (x_1(t), x_2(t), \ldots, x_N(t))$. 
We remove transients by deleting the first 10,000 data points and then downsample
the time series by considering only every 2nd value. Then, for
only 1,500 time points of the vector $\vec x(t)$ we calculate the RP and the above mentioned measures
DET, $1/L_{\max}$, RTE, and $D_{\Trans}$. We calculate this set of measures
for different system size $N \in \{10,\ldots, 50\}$ and repeat the calculation for
20 different initial conditions. For the line based RP measures DET and $1/L_{\max}$
we choose a minimal 
line length of two. We apply a Theiler window of length 20 (in units of
iteration steps) and a recurrence threshold such that the fraction of recurrences in the RP
is 10\% (and using the Euclidean norm). We estimated the size of the Theiler window by the 
auto-correlation time, which is in average 20. The choice of the fixed recurrence rate for the threshold selection
is justified by the increase of the state space dimension with growing $N$ that would 
require a rescaling of the recurrence threshold. By fixing the recurrence rate we can avoid 
this rescaling.

The inverse of the longest diagonal line $1/L_{\max}$ as well as the RTE reveal a similar alternating
variation with $N$ as $\lambda_{\max}$ (Fig.~\ref{fig_lyap}b,c). The Pearson correlation between
these two RP based measures and $\lambda_{\max}$ is 0.745 (for  $1/L_{\max}$) and 0.750 (for RTE).
The strong correlation even for the used rather short data segment suggests that these
RP based measures are good estimators for studying the divergence behavior of
high-dimensional systems.

\begin{figure*}[htbp]
\centering \includegraphics[width=\textwidth]{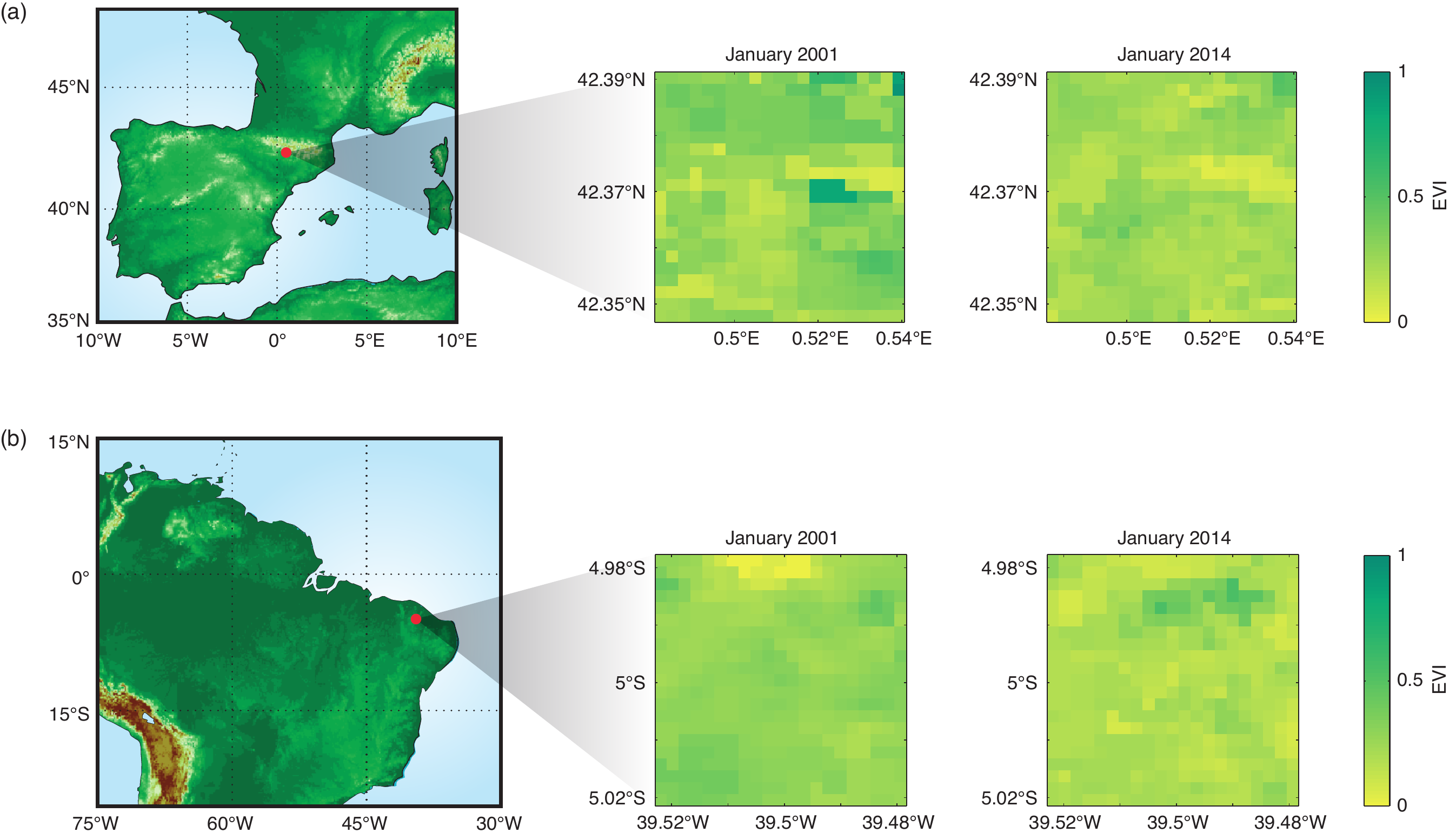} 
\caption{Geographical location and MODIS extended vegetation index (EVI) within the $5\,\times\,5$~km$^2$ 
subarea used for the analysis for the regions (a) NE Spain and 
(b) NE Brazil.}\label{fig_map_spain-brazil}
\end{figure*}

The DET measure varies between values of 0.94 and 1, indicating the deterministic
nature of the model (Fig.~\ref{fig_dim}b). During the periodic regimes, the DET shows maxima, whereas during the
chaotic regime, DET falls to lower values. The transitivity dimension $D_{\Trans}$
varies rather similar compared to the Kaplan-Yorke dimension $D_{\KY}$. It also shows the
upward trend with increasing $N$ (Fig.~\ref{fig_dim}c), but $D_{\KY}$ is in average 2.5 times higher than $D_{\Trans}$. 
The correlation of $D_{\KY}$ with DET and $D_{\Trans}$ is $-0.715$ and $0.723$, respectively.

The recurrence based measures are able to reveal the dynamics using very short time series of length 1,500, obtained from 
3,000 iterations, in comparison to the classic measures where 200,000 iterations and the differential equations had been necessary.
One explanation is that an RP compares the states at all time points with those at all other time points (i.e., an $N^2$ pair-wise
test), and that the measures are of statistical nature.  
The method will also work up to a certain level of noise. In the following we investigate the influence of noise by adding 
normally distributed random numbers to the time series. The standard deviation of the noise is
chosen relative to the mean amplitude $m_A$ of the time series and is varied between 0 (no noise) and
$0.5 \times m_A$. We find that even for a large portion of noise with standard deviation half of the mean amplitude,
the RQA measures distinguish clearly between the different dynamics (Fig.~\ref{fig_noise}). In particular, $D_{\Trans}$ shows
almost identical results for the considered noise levels (Fig.~\ref{fig_noise}(d)).
With increasing noise, DET and RTE decrease, and 
$1/L_{\max}$ increases. However, there are differences in their variations with respect to the noise
level. As the variation of DET and $1/L_{\max}$ increases for growing noise (Fig.~\ref{fig_noise}(b,d)), the variation 
for RTE decreases (Fig.~\ref{fig_noise}(c)).
It is remarkable that $D_{\Trans}$ is the measure with the lowest sensitivity on noise, whereas
$1/L_{\max}$ is less sensitive for low noise levels but becomes abruptly high sensitive for high 
noise levels (Fig.~\ref{fig_noise}(a)).
Nevertheless, these results suggest that the approach is quite robust even for higher level of observational
noise (at least for differentiating chaotic and periodic dynamics).

\section{Application on satellite time series imagery}

\begin{figure}[htp]
\centering \includegraphics[width=\columnwidth]{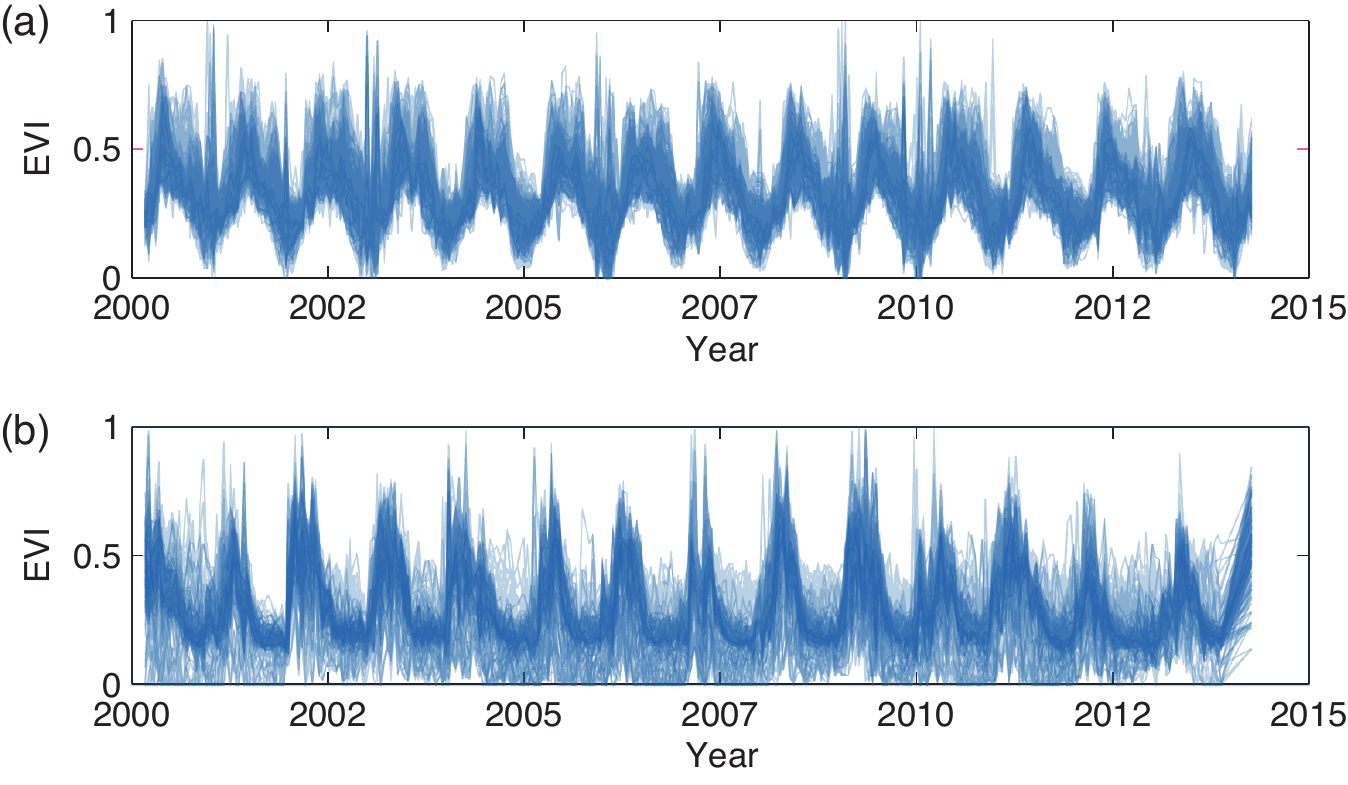} 
\caption{EVI time series of all pixels in the subareas as shown in Fig.~\ref{fig_map_spain-brazil}
for (a) NE Spain and (b) NE Brazil. The vegetation dynamics in NE Brazil appears clearly to be
more erratic in the temporal and spatial domain than in NE Spain.}\label{fig_modis_data}
\end{figure}

In order to illustrate the applicability of the proposed RP quantitative measures on spatially extended and
potentially high-dimensional real world data, we use MODIS satellite 
time series imagery of the extended vegetation index (EVI) 
of two test sites in NE Spain, centre coordinates 42.37$^\circ$N, 0.51$^\circ$E,
and NE Brazil, 5.00$^\circ$S, 39.50$^\circ$W (Fig.~\ref{fig_map_spain-brazil}). The test sites are characterized by differently complex 
vegetation dynamics both in the temporal (inter-annual and intra-annual) and spatial domain (Fig.~\ref{fig_modis_data}) as a result 
of diverse natural processes and human interactions \cite{barbosa2006,hill2008}. Thus, these sites are seen as ideal to study the 
usefulness of the proposed RP measures in order to objectively quantify and evaluate this complex behavior and decipher changes 
in vegetation cover dynamics related to land extensification/ intensification or climate change and drought. 
The subhumid Spanish test site shows a pronounced seasonal variation in precipitation and temperature 
with cold and dry winters and hot and stormy summers, whereas the Brazilian test site located in the so-called 
drought polygon is characterized by a semiarid climate with distinct dry and wet seasons and rainfall of high 
temporal and spatial irregularity. The Spanish test site has undergone severe land use changes during the last 
50 years with the abandoning of former agricultural areas and subsequent reforestation as well as setting 
aside of lands from agriculture promoted by the European Agricultural Policy \cite{lasanta2012}. The Brazilian 
test site has been more intensively occupied since 1985, when the Federal Government accomplished a land 
reform leading to the intensification of agricultural and livestock practices. A dense water surface reservoir 
network has been built in the last decades to mitigate water scarcity problems \cite{toledo2014}. 

The MODIS-Terra MOD13Q1 product used for this real world application is a 16-day composite image of the enhanced 
vegetation index (EVI) in a sinusoidal projection with a spatial resolution of 250~m. Global MODIS vegetation indices 
are designed to provide consistent spatial and temporal datasets used for global monitoring of vegetation 
conditions. The EVI is chosen 
since it minimizes canopy background variations and maintains sensitivity over dense vegetation.
We obtained 316 MOD13Q1 images for the period February 2000 to November 2013 for both the MODIS tiles h18v04 (Spain) and h14v09 (Brazil) from the Land Processes Distributed Active Archive Center (LP DAAC), located at the US Geological Survey (USGS) Earth Resources Observation and Science (EROS) Center (\texttt{lpdaac.usgs.gov}).

In both regions we consider subareas of $5\,\times\,5$~km$^2$ ($N = 441$ grid points) varying around the centre
point by 0.25$^{\circ}$ and within a range of $[-0.5^{\circ}\ 0.5^{\circ}]$ (resulting in 25 subareas for both regions).
That way, the subareas contain a mixture of land covers representative for the test sites. For calculating the
RP, we create the phase space vector $\vec x$ from the pixels of the satellite image subarea, i.e., $\vec x$ has
441 dimensions (not to be confused with the dimension of the dynamics).

\begin{figure}[htbp]
\centering \includegraphics[width=\columnwidth]{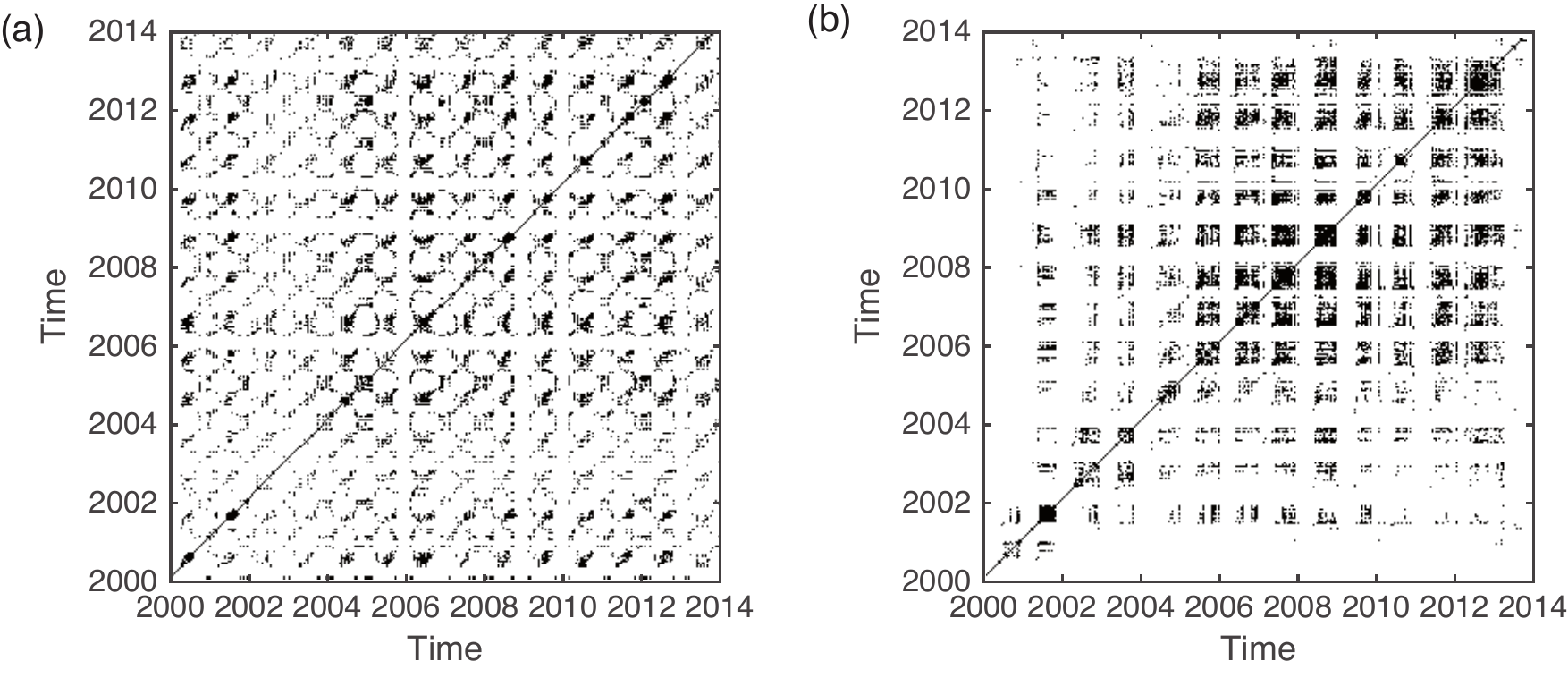} 
\caption{Recurrence plot of a $5\,\times\,5$~km$^2$ subarea of the EVI of test sites in (a) NE Spain and 
(b) NE Brazil. The selected subarea is situated at the centre point of the study region (see text).}\label{fig_rp_modis}
\end{figure}

\begin{figure}[htdp]
\centering \includegraphics[width=\columnwidth]{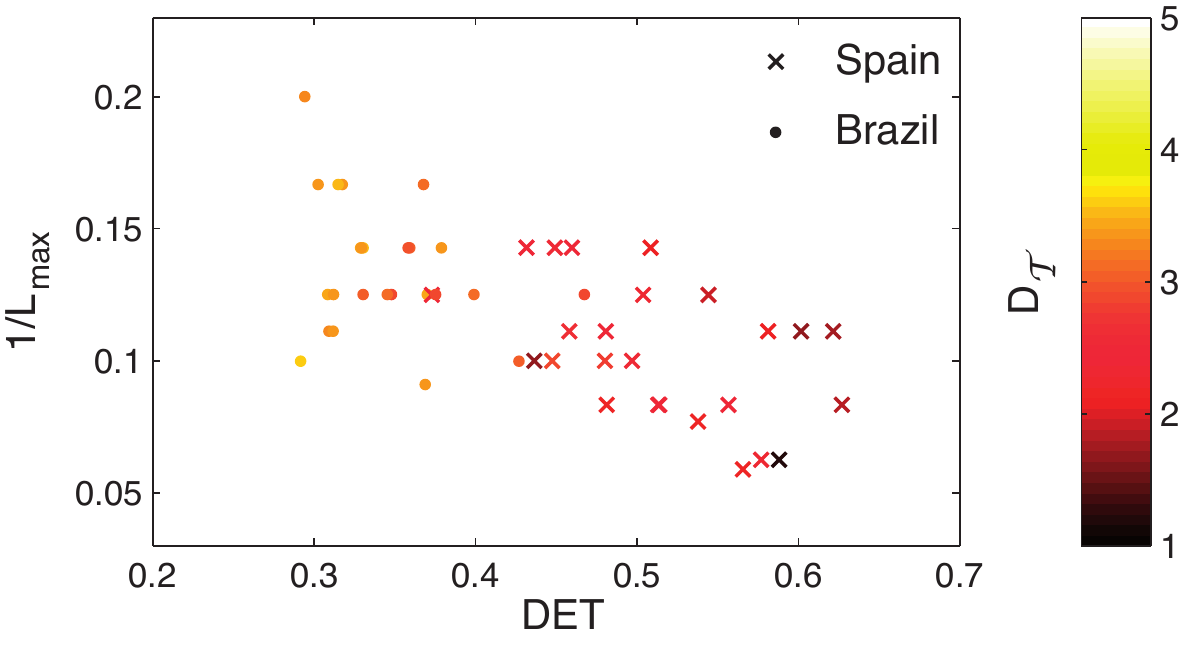} 
\caption{Recurrence quantification measures for the MODIS EVI data for different subareas around the study regions'
centre point.}\label{fig_rqa_modis}
\end{figure}

For both regions we visually find periodic patterns in the corresponding RPs, revealing mainly the seasonal 
variability (Fig.~\ref{fig_rp_modis}). The appearance of the periodic patterns differ for Spain (more line-like patterns)
and Brazil (more block-like patterns), indicating substantial differences in the spatial dynamics. The RP quantification
by the measures DET, $1/L_{\max}$, and $D_{\Trans}$ clearly reveals quantitative differences: in Brazil we find
a more erratic or chaotic spatio-temporal pattern than in Spain, indicated by lower DET and higher 
$1/L_{\max}$ as well as $D_{\Trans}$ for Brazil (Fig.~\ref{fig_rqa_modis}, Tab.~\ref{tab_rqa_modis}).
Although the considered subareas consist of information that is a mixed signal of several land cover classes,
the difference between Spain and Brazil is consistent for subareas of varying location. These results 
can be interpreted in such sense that the vegetation (or land use) dynamics in Brazil is probably less regulated
and less predictable than in Spain.

\begin{table}[htdp]
\caption{Median of recurrence quantification measures for the MODIS EVI data (standard deviation in brackets).}
\begin{center}
\begin{tabular}{lrr}
	&Spain	&Brazil\\
\hline
DET	&0.51 (0.07)	&0.33 (0.04)\\
1/$L_{\max}$	&0.10 (0.03)	&0.13 (0.03)\\
$D_{\Trans}$	&2.35 (0.45)	&3.85 (0.41)\\
\hline
\end{tabular}
\end{center}
\label{tab_rqa_modis}
\vspace{-14pt}
\end{table}

\section{Conclusion}
By using the Lorenz96 model as a prototypical example of spatially extended dynamics with large degrees of freedom, we have shown that
recurrence plot based analysis can be used to investigate high-dimensional dynamics from rather short time series and provides
insights in the fundamental features of the dynamics, comparable with the Kaplan-Yorke dimension or
the Lyapunov exponent. This study, thus, answers the hitherto open question, whether recurrence plots
and their quantification are suitable to study high-dimensional chaos. The more systematic study on the limits
of the used methods and the necessary length of time series in dependence on the degrees of freedom of the system
is a subject of future work. 

Moreover, by applying the 
method to MODIS satellite time series data we have demonstrated its suitability for the investigation of
extended spatio-temporal dynamics of real world processes. 
The recurrence analysis has indicated a clear difference in the spatio-temporal vegetation dynamics in a subhumid 
(Spain) and in a semiarid (Brazil) climate, where the first shows a more regular pattern, whereas the 
latter is characterized by a more irregular and less predictable behavior.

\section*{Acknowledgements}
We acknowledge support from the DFG and FAPESP (projects MA 4759/4-1 and IRTG 1740/TRP 2011/50151-0)
and from the Government of the Russian Federation (Agreement No.~14.Z50.31.0033).


\section*{References}

\bibliographystyle{elsarticle-num}

\bibliography{rp,others,mybibs}

\end{document}